\documentclass[aps,prd,twocolumn,groupedaddress,nofootinbib,
superscriptaddress,preprintnumbers]{revtex4-1}
\usepackage{epsfig,graphicx}
\usepackage{dcolumn}
\usepackage{bm}
\usepackage{amsfonts}
\usepackage{amssymb}
\usepackage{amsmath}

\usepackage{color}

\usepackage[hidelinks]{hyperref}

\usepackage{soul}

\begin{document}

\title{The Regular Ricci-Inverse Cosmology with Multiple Anticurvature Scalars} 

\author{Yicen Mou}
\email[]{myicen@hubu.edu.cn}
\affiliation{School of Physics, Hubei University, Wuhan, Hubei 430062, China}
\affiliation{Key Laboratory of Intelligent Sensing System and Security(Hubei University), Ministry of Education}


\begin{abstract}

We investigate the modified gravity in which the Lagrangian of gravity is a function of the trace of the $n$-th matrix power of Ricci tensor in a Friedmann–Lema\^{i}tre–Robertson–Walker(FLRW) spacetime. When $n$ is negative, the inverse of Ricci tensor, also called the anticurvature tensor, will be introduced. We design a new class of Ricci-inverse theory containing two anticurvature scalars and resulting to be free from the singularity problem.

\end{abstract}

\maketitle

{\bf Introduction.}
~Exploring the theory of gravity beyond the general relativity(GR) is a long-term collective activity that began more than one hundred years ago. The various motivations have give rise to many different forms of gravitational theories, e.g. Eddington's affine theory\cite{eddington1923}, Weyl's efforts for unifying gravitation and electromagnetism\cite{Weyl:1919fi}, the scalar-tensor theory by Brans and Dicke aimed to make the Newton's gravitational constant dynamical\cite{Brans:1961sx,Dicke:1961gz}, the theory with torsion in order to incorporate the quantum mechanical spin of elementary particles or in attempts to formulate gauge theories of gravity\cite{Hehl:1976kj} and many others. As time goes by, the modern precise cosmological observations hint at the existence of dark sector\cite{SupernovaSearchTeam:1998fmf,Bernal:2016gxb,GRAVITY:2021xju}, which prompts the creativity of theorists to find explanations in modified gravity if new particles beyond the Standard Model are excluded \textit{a priori}.

In order to have an observable effect at the present epoch, the coupling constant in the modified gravity should be of the order of powers---which depend on the physical dimension of coupling constant---of Hubble parameter $H_0$, this would lead to fine-tuning problem immediately\cite{Weinberg:1988cp,Sakstein:2019fmf}. Moreover, the new fields introduced do not have widely acknowledged relation to curvature, this is a departure from the geometrical view that has been held since Einstein. That's why besides those above, we have a huge number of theories based on the function of curvature tensor only\cite{Carroll:2004de,Capozziello:2011et,Glavan:2019inb,Lee:2020upe,Mustafa:2021mgh}, which can be regarded as the generalization of Einstein-Hilbert term.

In this paper, we focus on the Ricci curvature tensor only, deny the engagement of Riemann tensor. Since the Lagrangian should be a scalar, it's easily seen that they are actually the functions of the trace of Ricci tensor's matrix power after Talor/Laurent expasion. In this paper, we denote $R^{\mu}_{(1)\nu}$ as Ricci tensor and
\begin{equation}
	R^{\mu}_{(m)\alpha}R^{\alpha}_{(n)\nu}=R^{\mu}_{(m+n)\nu},\quad
	R_{n}=R^{\mu}_{(n)\mu}
\end{equation} 
Obviously $n=0$ case is the Kronecker delta. $n=-1$ gives the inverse of Ricci tensor, which is also called the anticurvature tensor. This is a complicated rational function of Ricci tensor after the Cramer's rule is used.

The most studied gravity in this category is of course the $f(R_{1})$ theory\cite{Sotiriou:2008rp,DeFelice:2010aj,Clifton:2011jh}, and to a lesser extent the $f(R_{1},R_{2})$ theory\cite{Noh:1998hd,Ohta:2018sze,Kaczmarek:2021psy}. We haven't cared about negative indices in quite a long time until the non-local operators were getting attention\cite{Deser:2007jk,Deser:2019lmm,Amendola:2020qho}. Theories contain anticurvature scalar(negative index $R_n$) are usually called the Ricci-inverse gravity, they have been extensively studied in the currently available literature in many aspects\cite{Do:2021fal,Jawad:2022eoj,deSouza:2023egd,Malik:2023hhn,Ahmed:2023lxz,Ahmed:2024jcb,Ahmed:2024urv,Ahmed:2024ezi,Ahmed:2024jsh,Moreira:2024mpf,Malik:2024zqr}.

In a spatially flat FLRW spacetime, the anticurvature scalars are rational functions of the Hubble parameter $H$ and its time derivation, while the positive index ones are power functions. This makes the Ricci-inverse gravity and positive index theory very different in nature. We are always fighting with the singularity in the Lagrangian and the evolution of the equations of motion when studying Ricci-inverse gravity. Due to the fact that $R^{0}_{(1)0}=0$ is the dividing line of cosmic acceleration/deceleration, there is a well known no-go theorem that any Lagrangian density $\mathcal{L}=R_{1}+\alpha R_{-1}^{l}$ cannot smoothly join a cosmic decelerated era with the current accelerated expansion of universe, no matter $l$ is positive or negative.

There are two ways to cover this singularity, $i)$ try constructing other type of function $f(R_{\pm1})$ in Lagrangian or $ii)$ make other $R_{n}$ involved. Since the denominator of $R^{-1}_{-2}$ is positive definite, it's a favored candidate. Recently, M.Scomparin gives a workable $f(R_{\pm1})$ model by proper construction \cite{Scomparin:2024poo}, from which the equation of motion(EoM) are unsurprisingly full of rational functions. I.Das and et al. choose the latter way, they analyses many models both in $f(R_{\pm1})$ and $f(R_{1},R_{-2})$ theory\cite{Das:2021onb}, the EoM is much more complicated. To our knowledge, there are few literature contains more than one anticurvature because of the redundant EoM\cite{deSouza:2024vqw,Ahmed:2024emk}, luckily the modern computer technology allows us hosting intermediate variables to make things easier. In this paper, we study the Ricci-inverse gravity containing both of $R_{-1}$ and $R_{-2}$ in FLRW spacetime without singularity, although our construction looks complex, the EoM can be rather simpler.

{\bf Specific Model.}
~We start our discussion with the modified Einstein
Hilbert action in a general model
\begin{equation}
	S=\frac{1}{2}\int d^{D}x\sqrt{-g}f(\{R_{n}\})+S_{matter}
\end{equation}
where $\{R_{n}\}$ is the set of all Ricci-power scalars the theory involves and $D$ is the dimension of spacetime. In this paper, the natural unit $8\pi G=1$ is always used. Define the helpful intermediate variables
\begin{equation}
	P^{\mu}_{\phantom{\mu}\nu}:=\sum_{n}n\frac{\partial f}{\partial R_{n}}R^{\mu}_{(n)\nu},\quad
	Q^{\mu}_{\phantom{\mu}\nu}:=R^{\mu}_{(-1)\lambda}P^{\lambda}_{\phantom{\mu}\nu}
\end{equation}
The EoM after variation with respect to the metric could be written in one line
\begin{equation}\label{eomf}
	-\frac{1}{2}fg_{\mu\nu}+P_{\mu\nu}-\frac{1}{2}(2Q^{\lambda}_{\phantom{\lambda}(\mu;\nu)\lambda}-Q^{\alpha\beta}_{\phantom{\alpha\beta};\alpha\beta}g_{\mu\nu}-\Box Q_{\mu\nu})=T_{\mu\nu}
\end{equation}
The semicolon here means the covariant derivative, i.e. $Q^{\alpha\beta}_{\phantom{\alpha\beta};\alpha\beta}=\nabla_{\beta}\nabla_{\alpha}Q^{\alpha\beta}$, and $\Box$ is the d'Alembert operator.

On the other hand, the spatially flat FLRW metric is
\begin{equation}
	ds^{2}=a^{2}(\eta)(-d\eta^{2}+d\vec{x}^{2})
\end{equation}
then, the maximally symmetric Ricci tensor is diagonal with
\begin{eqnarray}\label{xy}
	\left\{
	\begin{aligned}
		&X:=a^{2}R^{0}_{(1)0}=(D-1)\mathcal{H}'\\
		&Y:=\frac{a^{2}}{D-1}R^{i}_{(1)i}=\mathcal{H}'+(D-2)\mathcal{H}^{2}
	\end{aligned}
	\right.
\end{eqnarray}
\begin{equation}
	R_{n}=a^{-2n}[X^{n}+(D-1)Y^{n}]
\end{equation}
where a prime stands for $d/d\eta$ and $\mathcal{H}=a'/a$.

Notice that there is an algebraic identity
\begin{eqnarray}
	&&\frac{mx+ny-(m+n)\frac{mx^{-1}+ny^{-1}}{mx^{-2}+ny^{-2}}}{m+n-\frac{(mx^{-1}+ny^{-1})^{2}}{mx^{-2}+ny^{-2}}}=x+y
	\\
	&\Rightarrow&L_{n}:=\frac{R_{n}-D\frac{R_{-n}}{R_{-2n}}}{D-\frac{R^{2}_{-n}}{R_{-2n}}}=a^{-2n}[X^{n}+Y^{n}]
	\label{ln}
\end{eqnarray}

From the definition, there are 3 singular points for $L_{n}$: $X=0$, $Y=0$ and $X=Y$. The good thing is that they are all removable singularities, ensuring the Lagrangian made by the function of positive index $R_{n}$ and $L_{n}$ easily free from singularity.

As for the EoM, we take the variable $L_{1}$ as example, i.e.
\begin{equation}
	f=R_{1}+F(L_{1})
\end{equation}
Substituting it to \eqref{eomf} leads to ($D=4$ used)

\begin{eqnarray}
	\rho&=&3H^{2}-2(\xi-1)F_{L}H^{2}+\frac{1}{2}F
	\nonumber\\
	&&+8(2\xi^{2}-\xi-1)F_{LL}H^{4}+8F_{LL}H^{4}\xi_{\mathcal{N}}
	\label{friedmann}\\
	p&=&-(2\xi+1)H^{2}-\frac{1}{2}F
	\nonumber\\
	&&-\frac{8}{3}[3\xi(2\xi^{2}-\xi-1)+(7\xi-1)\xi_{\mathcal{N}}+\xi_{\mathcal{NN}}]F_{LL}H^{4}
	\nonumber\\
	&&-\frac{32}{3}(\xi_{\mathcal{N}}+2\xi^{2}-\xi-1)^{2}F_{LLL}H^{6}
	\label{pressure}
\end{eqnarray}
\begin{equation}
	T^{\mu\nu}_{\phantom{\mu\nu};\nu}=0
	\Rightarrow
	\rho'+3(1+w)\mathcal{H}\rho=0
	\label{conserve}
\end{equation}

where $\xi=\mathcal{H}'/\mathcal{H}^{2}$, $H=\mathcal{H}/a$, the subscripts $\mathcal{N}$ and $L$ stand for derivatives to $\ln a$ and to $L_{1}$ respectively. $w=p/\rho$ is the equation of state(EoS) parameter. Since the pressure relation \eqref{pressure} contains higher order derivative $\xi_{\mathcal{NN}}$, usually we choose the modified Friedmann equation \eqref{friedmann} together with the conservation equation \eqref{conserve} to form a complete set.

In the most frequent scenarios, the singularities in the cosmological evolution arise when the coefficient of $\xi_{N}$ hit its isolated zero/singular point, that could be controlled by $F_{LL}$ easily in our model. Furthermore, compare to the $f(R_{1})$ theory
\begin{eqnarray}\label{eomfr}
	f&=&R_{1}+\tilde{F}(R_{1})
	\nonumber\\
	\rho&=&3H^{2}-3\xi \tilde{F}_{R}H^{2}+\frac{1}{2}\tilde{F}
	\nonumber\\
	&&+36(\xi^{2}-1)\tilde{F}_{RR}H^{4}+18\tilde{F}_{RR}H^{4}\xi_{\mathcal{N}}
\end{eqnarray}

It's easy to see that \eqref{friedmann} and \eqref{eomfr} have a similar form, we can use the same method as in $f(R_{1})$ theory to study/cure the singularities in the evolution of FLRW background\cite{Nojiri:2008fk,Appleby:2009uf}. Specifically, when choosing the linear $L_{1}$ model, i.e. $F=\alpha L_{1}$ with a dimensionless constant $\alpha$, the $F_{LL}$ term is vanished and the EoM should be reduced to
\begin{eqnarray}\label{grcopy}
	\rho&=&3(1+\alpha)H^{2}
	\nonumber\\
	p&=&-(1+\alpha)(2\xi+1)H^{2}
\end{eqnarray}
It seems to be a copy of GR, the solution is nearly the same but with a shifted Hubble parameter
\begin{equation}
	H_{0}\rightarrow H_{\alpha}=\sqrt{1+\alpha}H_{0}
\end{equation}
Then, we get a regular cosmic evolution in the Ricci-inverse gravity with EoM as simple as in GR. Conversely speaking, from \eqref{grcopy} we have an additional component with the effective matter density $\rho_{eff}=-3\alpha H^{2}$ but without changing the EoS parameter, just as if the cold dark matter in a dust dominated universe.

{\bf Perturbations.}
~For simplicity, we consider a linear $L_{1}$ model with a Klein-Gordon(KG) field $\phi$.
\begin{equation}
	S=\frac{1}{2}\int d^{4}x\sqrt{-g}[R_{1}+\alpha L_{1}-(\partial\phi)^{2}-2V(\phi)]
\end{equation}

To get the perturbed expansion, we need to expanse the inverse matrix first. Decompose the Ricci tensor as follows
\begin{eqnarray}
	&&R^{\mu}_{(1)\nu}=a^{-2}(M^{\mu}_{\phantom{\mu}\nu}+N^{\mu}_{\phantom{\mu}\nu}),\quad
	M^{\mu}_{\phantom{\mu}\nu}=\textrm{diag}(X,Y\delta^{i}_{\phantom{i}j})
	\nonumber\\
	&&N^{0}_{\phantom{\mu}0}=0,\quad
	N^{0}_{\phantom{\mu}i}=a^{2}R^{0}_{(1)i},\quad
	N^{i}_{\phantom{\mu}0}=a^{2}R^{i}_{(1)0}
	\nonumber\\
	&&N^{i}_{\phantom{\mu}j}=a^{2}R^{i}_{(1)j}-Y\delta^{i}_{\phantom{i}j}
\end{eqnarray}
The traceless matrix $N^{\mu}_{\phantom{\mu}\nu}$ is perturbations, while the diagonal matrix $M^{\mu}_{\phantom{\mu}\nu}$ is a mixture of background and perturbation values. The inverse matrix could be written as a series
\begin{equation}
	R_{(-1)}=a^{2}(M^{-1}-M^{-1}NM^{-1}+M^{-1}NM^{-1}NM^{-1}+\cdots)
\end{equation}
Without confusion, we omit the indices in equation above to simplify the writings. Since $N$ is traceless, the expansion of $R_{n}$ up to order 2 is
\begin{eqnarray}\label{rn}
	R_{1}&=&a^{-2}\textrm{Tr}M=a^{-2}[X+(D-1)Y]
	\nonumber\\
	R_{-1}&=&a^{2}\textrm{Tr}[M^{-1}+NM^{-1}NM^{-2}]
	\nonumber\\
	&=&a^{2}[X^{-1}+(D-1)Y^{-1}
	\nonumber\\
	&&+\frac{x^{-1}+y^{-1}}{xy}N^{0}_{\phantom{0}i}N^{i}_{\phantom{i}0}
	+y^{-3}N^{i}_{\phantom{i}j}N^{j}_{\phantom{j}i}]
	\nonumber\\
	R_{-2}&=&a^{4}\textrm{Tr}[M^{-2}+NM^{-2}NM^{-2}+2NM^{-1}NM^{-3}]
	\nonumber\\
	&=&a^{4}[X^{-2}+(D-1)Y^{-2}
	\nonumber\\
	&&+\frac{2}{xy}(x^{-2}+x^{-1}y^{-1}+y^{-2})N^{0}_{\phantom{0}i}N^{i}_{\phantom{i}0}
	\nonumber\\
	&&+3y^{-4}N^{i}_{\phantom{i}j}N^{j}_{\phantom{j}i}]
\end{eqnarray}

$x,y$ here are the background values of $X,Y$ in \eqref{xy} respectively. Substituting \eqref{rn} into \eqref{ln} we get
\begin{eqnarray}
	L_{1}&=&a^{-2}\left[X+Y+\frac{D-2}{D-1}\frac{N^{0}_{\phantom{0}i}N^{i}_{\phantom{i}0}}{x-y}-\frac{J}{x-y}N^{i}_{\phantom{i}j}N^{j}_{\phantom{j}i}\right]
	\nonumber\\
	J&\equiv&\frac{x}{y}\left(\frac{x}{y}+\frac{2}{D-1}\right)
\end{eqnarray}

From now on we always have $D=4$. The EoM connects parameters $x-y$ and $x/y$ to KG field $\phi$
\begin{eqnarray}
	\left\{
	\begin{aligned}
		&(1+\alpha)(-x+3y)=\phi^{\prime2}+2a^{2}V
		\nonumber\\
		&(1+\alpha)(-x-y)=\phi^{\prime2}-2a^{2}V
	\end{aligned}
	\right.
	\\
	\Rightarrow
	\left\{
	\begin{aligned}
		&x-y=2(\mathcal{H}'-\mathcal{H}^{2})=-\phi^{\prime2}/(1+\alpha)
		\\
		&\frac{x}{y}=\frac{3\mathcal{H}'}{\mathcal{H}'+2\mathcal{H}^{2}}=1-\frac{\phi^{\prime2}}{a^{2}V}
		=\frac{3w+1}{w-1}
	\end{aligned}
	\right.
\end{eqnarray}

After the doing so, we find that the singularities may show up again. Comparing to the model $f=R_{1}+\frac{\alpha R_{1}}{\beta+R_{-1}R_{1}}$ in literature slightly earlier\cite{Scomparin:2024poo}, there's a possible singular point in the dividing line of cosmic acceleration and deceleration $x=0$, we move this risk to de Sitter phase $x=y$. The other singular point $y=0$ corresponds to $w=1$, we do not normally consider such a large EoS parameter.

One of the key observables is the cosmological scalar perturbation, which is essentially due to the matter sector and the scalar polarization of metric fluctuation it induces. We define the scalar perturbation on the metric and matter as follows,
\begin{eqnarray}
	&g_{00}=-a^{2}(1+2A),\quad g_{0i}=a^{2}\partial_{i}B
	\nonumber\\
	&g_{ij}=a^{2}[(1-2\psi)\delta_{ij}+\partial_{i}\partial_{j}E]
	\nonumber\\
	&\phi(\eta,\vec{x})=\phi(\eta)+\delta\phi(\eta,\vec{x})
\end{eqnarray}

Note that the theory possesses full spacetime diffeomorphisms, and therefore we can safely remove $\delta\phi$ and $E$ by performing a proper coordinate transformation. After using the EoM and doing some tedious integration by parts, the 2nd order action in Fourier space is
\begin{eqnarray}\label{scalars2}
	S_{2}&=&\frac{1+\alpha}{2}\int d\eta d^{3}ka^{2}\{
	-2yA^{2}-6\psi^{\prime2}-12\mathcal{H}A\psi'
	\nonumber\\
	&&+4k^{2}B(\mathcal{H}A+\psi')-2v_{g}^{2}k^{2}(2A\psi-\psi^{2})
	\nonumber\\
	&&+\frac{2\alpha k^{2}}{3\phi^{\prime2}}[4(\mathcal{H}A+\psi')^{2}+Jk^{2}(B'+2\mathcal{H}B+A-\psi)^{2}]
	\}
	\nonumber\\
	v_{g}^{2}&\equiv&\frac{1+\frac{\alpha}{3}}{1+\alpha}
\end{eqnarray}

Different from GR, $B$ is no longer a Lagrangian multiplier at this time while $A$ stays the same, meaning that only one constraint equation remains and an extra degree of freedom(DoF) arises, like many other gravitational theory with higher order derivative do.
\begin{eqnarray}
	&&\left[-2y+\frac{2\alpha k^{2}}{3\phi^{\prime2}}(4\mathcal{H}^{2}+Jk^{2})\right]A
	=6\mathcal{H}\psi'-2k^{2}\mathcal{H}B
	\nonumber\\
	&&+2v_{g}^{2}k^{2}\psi
	-\frac{2\alpha k^{2}}{3\phi^{\prime2}}\left[4\mathcal{H}\psi'+Jk^{2}(B'+2\mathcal{H}B-\psi)\right]
	\nonumber\\
\end{eqnarray}
Solving this constraint for $A$ would place $k$ on denominator, which reflects the non-local aspect for the Ricci-inverse theory

Since we do not intend to do an overly detailed analysis in this paper, we will not substitute the constraint equation here and list the two EoM in fourth order derivative for the scalar perturbation, but pointing out that if $J$ trended to zero and $a^{2}J/\phi^{\prime2}$ was nearly time independent, $B$ will give a constraint again, leaving
\begin{equation}\label{pgr}
	S_{2}\simeq\frac{1+\alpha}{2}\int d\eta d^{3}k\frac{a^{2}\phi^{\prime2}}{\mathcal{H}^{2}}(\zeta^{\prime2}-v_{g}^{2}k^{2}\zeta^{2})
\end{equation}

The curvature perturbation of hypersurfaces with homogeneous $\phi$ field $\zeta=-\psi-\mathcal{H}\delta\phi/\phi'$ is coincident with $-\psi$ under our gauge condition. Here the scalar DoF are merged into one, the only difference from GR now is the group velocity $v_{g}$, we'll call it a pseudo GR state in the following. The energy condition \eqref{grcopy} constraints $\alpha>-1$, so keeping $v_{g}^{2}$ positive definite. $J=0$ could happen when $w=-1/3$ or $w=-1/11$, corresponding to the redshift $z=0.67$ or $z=1.8$ in a $\Lambda$CDM model, these are the key periods for structure formation.

As for the de Sitter limit, we make an ansatz that since the background evolution is always regular, the perturbation should not blow up. This requires that the square bracketed parts in \eqref{scalars2} get vanished at this point, those are what we have in GR. At such a stage, $B$ also goes back to be non-dynamical, making the pseudo GR \eqref{pgr} appear again.

Another important observable in the FLRW universe is the transverse and traceless part of the metric fluctuation---the tensor modes---which we define as
\begin{equation}
	g_{ij}=a^{2}(\delta_{ij}+2h_{ij})
\end{equation}
where $h_{ij}$ satisfies $k_{i}h_{ij}=0$ and $h_{ii}=0$. At linear level the tensor modes are gauge invariant. Their 2nd action reads
\begin{eqnarray}\label{tensors2}
	S_{2}&=&\frac{1+\alpha}{2}\int d\eta d^{3}ka^{2}[h^{\prime2}_{ij}-v_{g}^{2}k^{2}h_{ij}^{2}
	\nonumber\\
	&&+\frac{\alpha J}{\phi^{\prime2}}(h_{ij}''+2\mathcal{H}h_{ij}'+k^{2}h_{ij})^{2}]
\end{eqnarray}

Unsurprisingly the EoM for tensor modes will also be a 4th order differential equation. When scalar modes get the pseudo GR phase in the $J$ vanished limit, we have pseudo GR for tensor modes
\begin{equation}\label{pgrtensor}
	S_{2}\simeq\frac{1+\alpha}{2}\int d\eta d^{3}ka^{2}(h^{\prime2}_{ij}-v_{g}^{2}k^{2}h_{ij}^{2})
\end{equation}
It also differs from GR with only a parameter $v_{g}^{2}$. Things are totally changed in de Sitter limit, even we make the same ansatz as in scalar modes that terms proportional to $\phi^{\prime-2}$ should trend to zero, i.e.
\begin{equation}\label{tensor1}
	h_{ij}''+2\mathcal{H}h_{ij}'+k^{2}h_{ij}=0
\end{equation}
meaning that we may get \eqref{pgrtensor} again, which leads to another EoM
\begin{equation}\label{tensor2}
	h_{ij}''+2\mathcal{H}h_{ij}'+v_{g}^{2}k^{2}h_{ij}=0
\end{equation}

The conflict between \eqref{tensor1} and \eqref{tensor2} shows that there cannot be tensor mode fluctuations for de Sitter limit in our theory, or the 2nd line of \eqref{tensors2} should be finite. We leave the detailed analyses for future work.

{\bf Conclusion.}
~Gravitational theories that depend only on Ricci tensor are widespread in modern researches, theories involves the negative power of Ricci tensor are also called the Ricci-inverse gravity. These theories have been plagued by the singularity since its inception, and previous attempts to solve the issue have either had the EoM filled with rational functions and their composition, or have been discouraged by the sheer number of terms involved in EoM. In the present paper we design a class of variable $L_{n}$, which has the same physical dimension of $R_{n}$ and does not contain rational function in result. We analyses the influence of taking $L_{1}$ into the Lagrangian of gravity. Although $L_1$ involves two anticurvature scalar $R_{-1}$ and $R_{-2}$, the EoM is comparable with $f(R_1)$ theory by the help of unified EoM \eqref{eomf} for both of the positive and negative power of Ricci tensor. Accidentally, the linear $L_1$ model is even more likely to give results as simple as GR.

At the perturbation level, even a linear $L_1$ model has to show its fourth-order gravity nature. Furthermore, the singular risk may come back. But $x=0$, the dividing line of cosmic acceleration and deceleration, is still a removable singular point, meaning that our theory is safe both for background and fluctuations as long as the EoS parameter $w$ ranges between -1 and 1. We also demonstrate the possibility that when $x=0$ or $x=y$, the extra scalar DoF may disappear, making the evolution differs from what GR predict with only the group velocity. For tensor modes, the behavior is similar with scalar modes at $x=0$, but in de Sitter phase more detailed analyses is needed.


\end{document}